# Supporting tangible multi-factor key exchange in households

**Thomas Lodge, Sameh Zakhary, and Derek McAuley**

Horizon Centre, University of Nottingham, Nottingham Geospatial Building, Triumph Rd, Nottingham NG7 2TU, United Kingdom

A common approach to securing end-to-end connectivity between devices on the Internet is to utilise a cloud-based intermediary. With this reliance upon a third-party comes a set of security and privacy concerns that are difficult to mitigate. A promising new protocol, Wireguard, dispenses with the middleman to provide secure peer-to-peer communication. However, support for initial key exchange falls outside Wireguard's scope, making it potentially vulnerable to insecure out-of-band key exchange. The design of secure and usable key exchange methods is challenging, not least in domestic spaces, as they're often characterised by technically naïve users in multi-occupancy environments, making them susceptible to insider and passer-by attacks (i.e.: theft, observation attacks, relay and impersonation attacks). We describe and present the results from a design ideation study that probes the use of tangible, multi-factor approaches for securing key exchange in domestic spaces. The study suggests that a home's semi-fixed features (e.g.: lamps, shelves, chairs) can be instrumented to support a promising three-factor authentication approach ('*what you have*, *what you know* and *where you are*') to enable key exchange solutions that are *i.* more secure than commonly used naïve approaches and *ii.* desirable for end users.

## I.　　INTRODUCTION

There is a real and increasing need for householders to communicate securely with the IoT devices in their homes, even if not physically present within them (e.g., setting heating, accessing CCTV, communicating with couriers and so on). The predominant approach to securing end-to-end communication between devices on the Internet is to utilise a 'man-in-the-middle': a trusted (typically cloud-based) third party that handles rendezvous and communication management. There are, however, inherent performance and security weaknesses with this approach; the third party can become a bottleneck and is susceptible to traffic analysis and DDOS attacks. Moreover, reliance on third parties brings a set of privacy compliance headaches; as stewards of data, third parties will be subject to GDPR and the hefty penalties that arise from non-compliance.

Recent approaches, such as Wireguard, offer a solution that removes the dependency on a middleman; instead, supporting direct encrypted communication between peers. Wireguard relies on, but does not specify, out-of-band key exchange to establish 'permanent' keys, obviating the requirement for per-session login. Without adequate support, the mechanism employed to facilitate key exchange will be open to attack if users choose convenient, yet insecure approaches to exchange keys (such as email, usb sticks and so forth).

Secure, usable key exchange mechanisms are challenging to design and implement. However, by making *physical* interactions a requirement of key exchange, we can mitigate a whole class of online attacks. In designing for households, any key-exchange interactions must be designed with the assumption that end-users will have little to no networking experience and that the environment in which they take place is multi-occupancy. We therefore must assume the possibility that the system may be opportunistically subverted within the household by guests, roommates, passers-by or family. This necessitates multi-factor authentication approaches (MFA) that can draw upon secrets, physical access and ownership in order to limit attacks such as brute force, observation, impersonation or denial of service.

There is a beguiling array of multifactor authentication approaches proposed in research and deployed in the real world. Though we provide a more in-depth categorisation and critique later, it's clear that good multi-factor systems are hard to design, both in terms of security and user acceptance. With regard to user acceptance of MFA, FIDO[1], a large open industry alliance states simply and bluntly, that: "*consumers don't like the user experience*". Moreover, a large body of research, (Cristofaro, 2013), (Krol, 2015) (Weir, 2010) show that the use of multiple authenticators in the execution of a multifactor authentication protocol will often negatively affect user experience, which can, in turn, compromise security.

## II.    BACKGROUND

**Authentication approaches**

There are two parts to key exchange i. establishing authority (authentication and authorisation) and ii. key transfer over a secure channel. This section provides an overview of authentication approaches that bear relevance to our own work. To help locate the design choices we make in related literature, we orient our summary around the classical three categories of authentication (*what you know, what you have and what you are*) but extend it to include *where you are.*

*What you know* authenticates a user based on a shared secret; the most common example is text passwords, though graphical passwords (Belk, 2019) gestures (Kong, 2019), patterns (Meng, 2019) and interactional tasks (Leung, 2018) also fit within this category. Knowledge based schemes may be judged against two competing dimensions: usability and security. Security is most often improved by increasing complexity (longer/more complex passwords, gestures, tasks), though this can be to the detriment of usability. Nevertheless, *what you know* schemes have desirable properties; they're portable, well understood, shareable and often more straightforward to implement than alternative approaches (Bonneau, 2012)

*What you have* delegates authority to physical items (physical in the majority of cases; an email account one of a few exceptions, which can also be seen as what you have), that is, access or ownership becomes the proxy for identity and authority. We distinguish between two subtypes: *what you carry*, and *what you own*; each employ distinctly different interactions. *What you carry* requires that the authenticating item is mobile and carried on-person; it includes all types of wearable computing (glasses, smartwatches) (Stajano, 2011) as well as tokens (e.g. on keyrings), physical unclonable functions (PUF) (Gao, 2020) and smartphones. *What you carry* authentication approaches can be simple, they do not require users to commit anything to memory and users can use simple activities to improve security (e.g., hiding a token in the biscuit tin). Though the literature is rich with *what you carry* approaches (Mare, 2019) they all share common weaknesses: the need to carry and safeguard important objects and the risk of theft, loss and damage. Although leveraging smartphones as the 'carried' item makes good sense from a user adoption and usability perspective (given that people carry smartphones as a matter of course), it can undermine the principle of MFA, whose security relies upon independence between each of the authenticating channels; in cases where all channels are on the same device, compromising the device will compromise all channels (Hagalisletto, 2017).

*What you own*, leverages static or semi-fixed objects that inhabit the home or office (e.g. a lamp, shelves, a chair) and so rely on users having physical access to bespoke authenticating hardware in a semi-private space. Because there are few examples in the literature, our aim is to investigate whether this is a promising design space, i.e. one that furnishes common household objects with authenticating capabilities.

---
[1] https://fidoalliance.org/overview/

*What you are* refers principally to biometric profiling, that is, a person's unique characteristics provide the basis of authentication: fingerprints, retinal scan, voice, facial features (Bhattacharyya, 2019). This approach can, if properly implemented, offer strong authentication guarantees and seamless interactions; users neither need to carry dedicated objects or memorise secrets. However, biometric authentication can feel 'creepy'(McWhorter, 2021) and requires that sensitive data are stored somewhere, with all the privacy implications this entails. Moreover, there are unresolved, difficult problems in establishing *liveness* (Akhtar, 2015) i.e., that the biometric details used to authenticate actually come from a live person (rather than, for example a photo or recording).

Finally, *where you are* covers the wide range of solutions that utilise location, either absolute or relative (proxemic), to infer legitimacy. Solutions range from utilising short-range connectivity (RFID, line-of-sight), which will exclude anyone beyond a particular boundary, to using locally unique stable features of the environment (signal strength (Varshavsky, 2007), audio co-analysis (Shrestha, 2018), channel jamming (I.Martinovic, 2009) time of flight distance bounding (Mauw, 2018), to tracking a user's position within space (geolocation (Zhang, 2012)) to requiring interactions that can only be performed locally (e.g. Wifi Protected Setup[2] or QRCodes (Eminagaoglu, 2014)).

**Multi factor approaches**

Figure 1 provides an overview of four dimensions of authentication. The application of MFA varies across the services and domains that need to authenticate users; many cloud services (e.g: Gmail. Github, Twilio, Digitalocean) have converged on a 2-factor authentication approach that combines passwords (*what you know)* with *what you have* (email address, SMS, authentication apps), whereas online banking systems are more fragmented; relying upon a plethora of proprietary systems that range in both the types of multifactor channels employed and the services and processes that require MFA (e.g.: login, transferring funds etc.) (Sinigaglia, 2020). Barclays, Nationwide and others, for example, have employed a standalone security device (variously named: PIN Sentry, Card reader) that requires a bank card and customer PIN to generate a one-time code (*what you have* and *what you know*). Some banks (Barclays[3], HSBC[4], Halifax[5]) offer a single factor telephone banking service that identifies customers through voice recognition (*what you are*), though there are documented issues of spoofing[6] and manipulated input attacks (Y.Xie, 2020)(Abdullah, Hadi, 2020). Many banks now offer a mobile app, which leverages a smartphone's biometric capabilities (such as fingerprint or face recognition) (*what you have* and *what you are*). Such is the range of implemented solutions that a variety of initiatives have been proposed to push for standardisation of protocols (FIDO[7] and OAuth

---

[2] Alliance, W.F., 2007. Wi-fi protected setup specification. *WiFi Alliance Document*, 23.

[3] https://www.wired.co.uk/article/barclays-voice-security-telephone-banking
[4] https://engageemployee.com/hsbc-first-bank-to-use-voice-recognition-for-telephone-banking-customers/
[5] https://www.halifax.co.uk/contactus/call-us/voice-id/voice-id-faqs.html
[6] https://www.bbc.co.uk/news/technology-39965545
[7] https://fidoalliance.org/about/overview/

Authentication[8]) and to push for regulations and guidelines (European Banking Authority[9], NIST[10], PCI[11], Centrify[12], Gemalto[13], PingIndentity[14]).

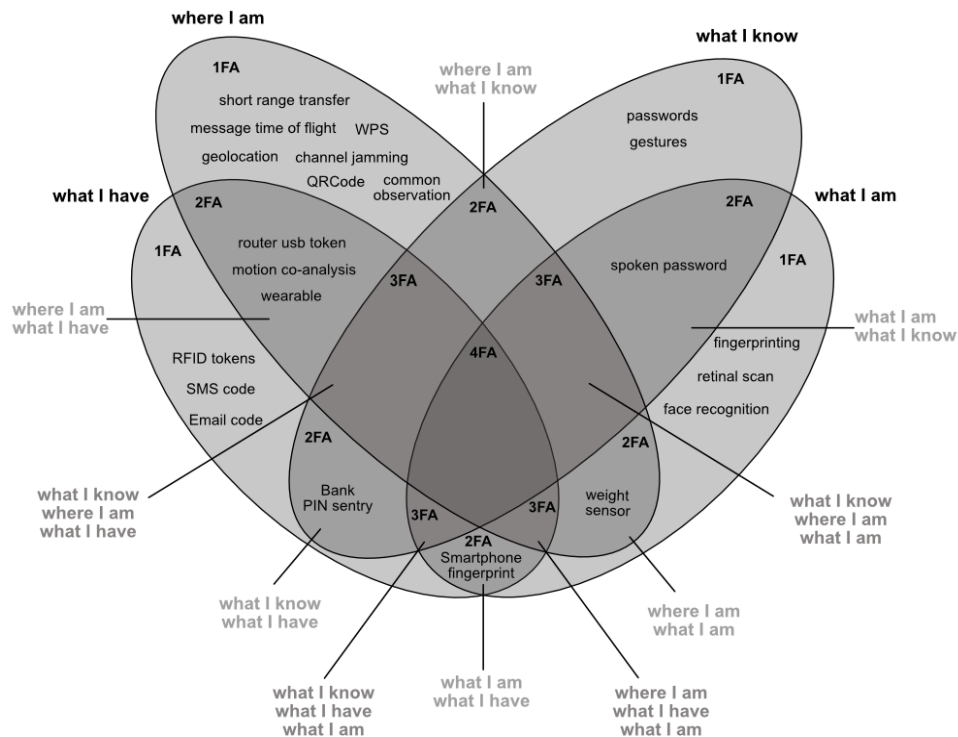

Figure 1: Four dimensions of multifactor authentication

Domestic IoT, perhaps in part due to its infancy and in part due to unbridled fragmentation across hardware, software, protocols, cost and application domains, offer multifactor authentication as the exception rather than rule. Most approaches that have been proposed that tackle the particular quirks of IoT in domestic environments (multi-modal interactions, interface constraints, hardware/performance constraints etc.) currently only exist in research (Ometov, 2019).

---

[8] https://openauthentication.org/
[9] https://eur-lex.europa.eu/legal-content/EN/TXT/PDF/?uri=CELEX:32018R0389&from=EN
[10] https://pages.nist.gov/800-63-3/
[11] https://www.pcisecuritystandards.org/pdfs/Multi-Factor-Authentication-Guidance-v1.pdf
[12] https://www.centrify.com/media/3403844/bpb-best-practices-for-multi-factor-authentication.pdf
[13] http://www2.gemalto.com/email/2011/authsomethingstronger/whitepaper/Authentication_Best_Practices_WP(EN)_A4_web.pdf.
[14] https://www.pingidentity.com/content/dam/ping-6-2-assets/Assets/white-papers/en/mfa-best-practices-securing-modern-digital-enterprise-3001.pdf?id=b6322a80-f285-11e3-ac10-0800200c9a66.

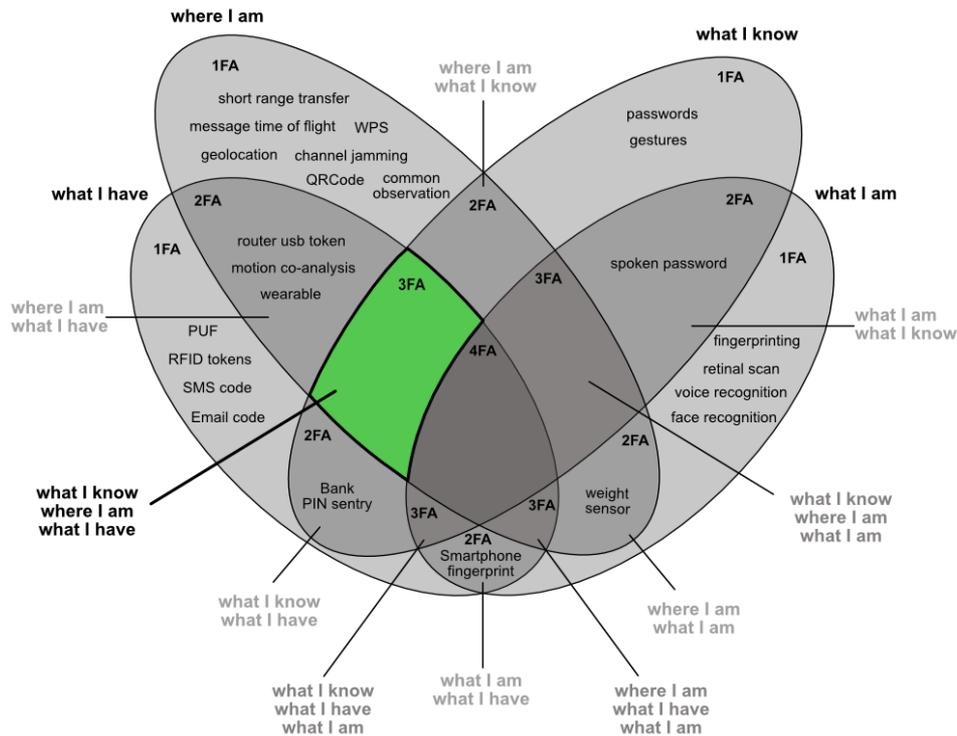

Figure 2: Design space: three factor authentication using semi-fixed features

There are three features of the user base in a household setting that must be designed for in any solutions that we consider:

1. Multi-user. Multiple users will need engage with the authentication system.
2. Non-technical users. Emphasizing the need for approaches that are usable and legible to the lay person.
3. Untrusted users. We have to assume 'attacker within', i.e. that visitors (and even fellow householders, passers-by) are untrusted.

We posit that there may be solution space in domestic IoT environments which leverages three dimensions of authentication (what you know, what you have, where you are) to offer solutions that are *both* secure and desirable for end users (Figure. 2)

### III. METHODOLOGY

The purpose of our study is to explore the merits and limitations of a *potential* solution space (rather than evaluation of a final solution) that utilises fixed features in a household as part of a multifactor authentication scheme. We adopted a 'sketching and ideation' approach, akin to a focus group study, to gather feedback on a range of alternative solutions. Each of our solutions share a set of common characteristics:

1. Technically feasible.
2. Tangible interactions with domestic fixtures and fittings. We presume that where users may resist 'yet another bit of hardware' they may be more receptive to the appropriation and extension of existing fixtures and fittings (what you have).

3. Constrained by location. Interactions are with semi-fixed features in the home (where you are).
4. Their secrets (what you know) are hidden in plain sight; expressed as orientations, locations, patterns or arrangements.

Each of the proposed solutions presume the home router is the VPN endpoint; that there is an out-of-the box setup that has a Wireguard endpoint, opened ports and static IP or dynamic DNS pre-configured only to permit incoming authenticated VPN traffic on specific ports. We also assume that the router will natively support any technology employed to communicate between the semi-fixed objects and router (e.g.: radio, line-of-sight, nearfield).

To familiarise users with potential solutions and to help imagine their use in their own homes, we created plausible product designs and renderings alongside descriptions and storyboards (to summarise proposed interactions with the system). The details of interaction were kept deliberately sparse, to encourage users to question details and raise concerns (Figure 3). We presented nine different solutions; each chosen to reflect distinctly different modes of interaction and aesthetic. All solutions are related by having a primary household purpose.

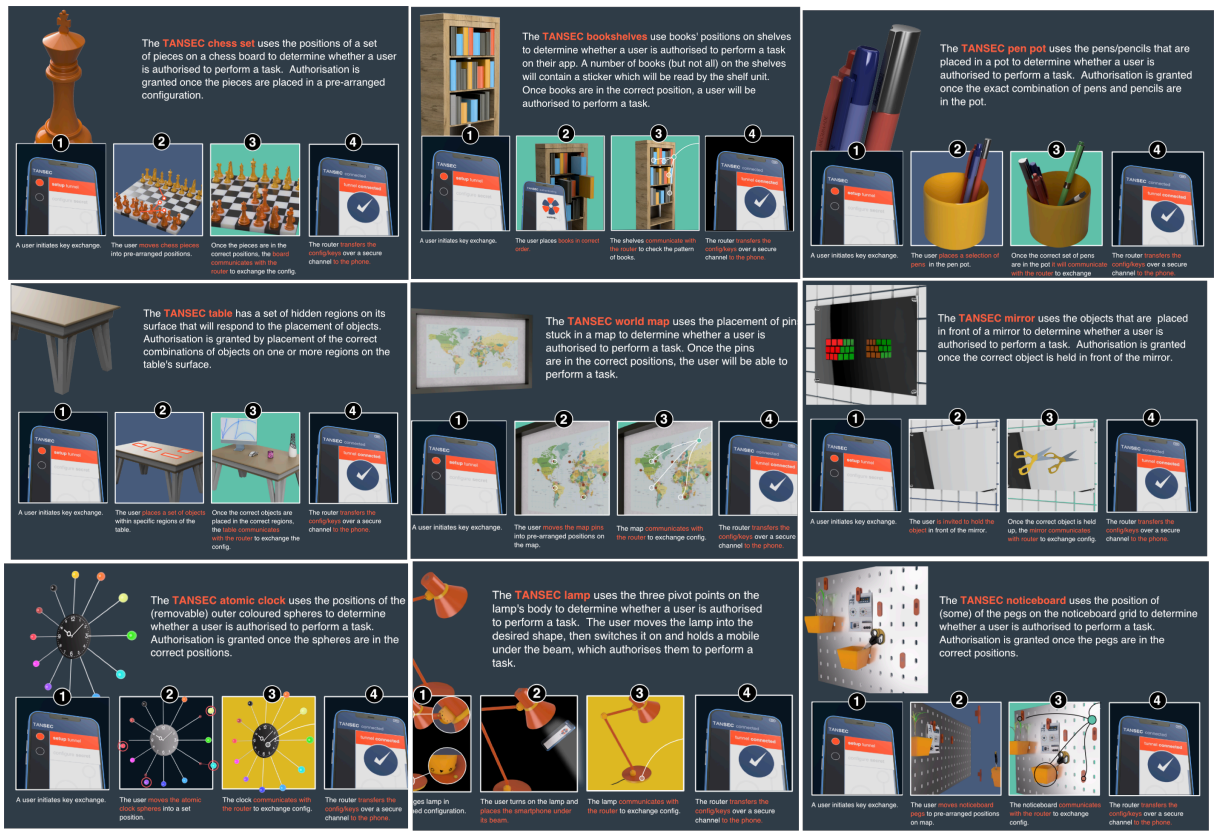

**Figure 3: Solution sketches and storyboards**

Each design was accompanied by a feedback questionnaire which was used to establish its: *i.* intelligibility (how easy it was to comprehend), *ii.* desirability (would users actually want/use it), *iii.* complexity (of interactions), *iv* perceived security. The questionnaire also required users to rank each sketch and to provide (unstructured) feedback on the overall response to solutions.

## Brief description of designs

| | |
|---|---|
| 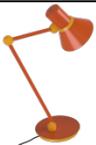 | **Lamp**<br>Uses the three pivot points on a lamp's body to determine whether a user is authorised to perform a task. The user moves the lamp into a desired shape then switches it on and holds a mobile under the beam which authorises them to perform a task. |
| 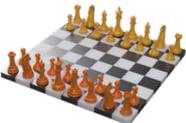 | **Chessboard**<br>Uses the position of a set of chess pieces on a chessboard to determine whether a user is authorised to perform a task. Authorisation is granted once the chess pieces are placed in a pre-arranged position. |
| 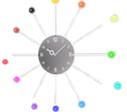 | **Atomic clock**<br>Uses the positions of the (removable) outer coloured spheres to determine whether a user is authorised to perform a task. Authorisation is granted once the spheres are in the correct positions. |
| 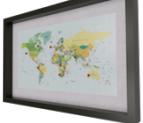 | **World map**<br>Uses the placement of pins stuck in a map to determine whether a user is authorised to perform a task. Once the pins are in the correct position the user will be authorised to perform a task. |
| 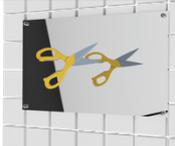 | **Mirror**<br>The mirror uses objects that are placed in front of the mirror to determine whether a user is authorised to perform a task. Authorisation is granted once the correct object is placed in front of the mirror. |
| 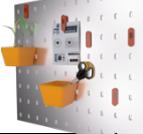 | **Noticeboard**<br>Uses the position of (some) of the pegs in the noticeboard grid to determine whether a user is authorised to perform a task. Authorisation is granted once the correct pegs are in the correct positions. |
| 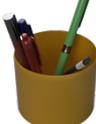 | **Pen pot**<br>Uses the pens/pencils that are placed in a pot to determine whether a user is authorised to perform a task. Authorisation is granted once the exact combination of pens and pencils are in the pot. |
| 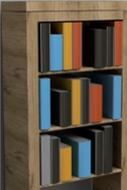 | **Shelves**<br>Uses books' positions on shelves to determine whether a user is authorised to perform a task. A number of books (but not all of them) will contain a sticker which will be read by the shelf unit. Once books are in the correct position, users will be authorised to perform a task. |
| 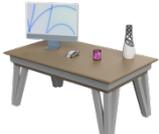 | **Table**<br>The table has a set of hidden regions on its surface that will respond to the placement of objects. Authorisation is granted by placement of the correct combination of objects on one or more regions on the table's surface. |

**Table one: brief description of designs**

## Results

We recorded participants' security expertise and household makeup to provide a little background context to the results. We received 28 responses to our survey, with an almost even split between respondents with 'good' or 'professional' expertise (54%) and those with little or no expertise (46%). There was an even split between households with one or two occupants (50%) (i.e. couples or lone households) and shared/family households (50%) see table two.

| Security knowledge: | Household makeup |
|---|---|
| good: 7, | family: 13, |
| little: 12, | couple: 11, |
| professional: 8, | shared: 1, |
| none: 1 | lone: 3 |

**Table two: respondent makeup**

## IV. RESULTS

We begin this section with a (quantitative) summary of results to assess the most popular solutions amongst different classes of respondents. We follow with a deeper dive into the security and usability feedback to: *i.* investigate the viability of this solution space, and *ii.* set out the criteria that must be considered when designing home based authentication solutions.

Our first concern was whether the solution space presented a viable area to explore, or whether the overall approach would be rejected by respondents. Four (14%) respondents stated that they would not use *any* of the solutions (i.e. in response to the question: "Would you ever consider using this system?"). Most users (75%) had different reactions to each solution (i.e. that they would consider using some, but not others). Three users (11%) stated that they would consider using *all* of the solutions in their houses. 61% of all respondents would consider using the two most popular solutions, whereas for the remaining seven solutions, fewer than half respondents would consider using them (see table three)

| Solution | % that would use it (all respondents) | % that would use it (security good /pro) | % that would use it (security little/none) | % that would use it (family / shared) | % that would use it (lone/couple) |
|---|---|---|---|---|---|
| map | 61 | 53 | 69 | 57 | 64 |
| mirror | 61 | 73 | 46 | 50 | 71 |
| lamp | 43 | 47 | 38 | 43 | 43 |
| chess | 39 | 40 | 38 | 29 | 50 |
| noticeboard | 39 | 40 | 38 | 36 | 43 |
| table | 35 | 33 | 38 | 29 | 43 |
| shelves | 39 | 33 | 38 | 21 | 50 |
| clock | 25 | 27 | 30 | 21 | 29 |
| penpot | 18 | 20 | 8 | 7 | 29 |

**Table three: respondents who would use the solutions**

Users were asked to provide an overall ranking on a 5-point Likert scale (0, I hate it – 4 , I love it) for each of the solutions. The map achieves the highest overall score across all categories; there is a little re-ordering around the lower ranked solutions.

| Mean Score | (Security: good/pro) | (Security: little/none) | Household: (family/shared) | Household: (lone/couple) |
|---|---|---|---|---|
| 1 Map | 1 Map | 1 Map | 1 Map | 1 Map |
| 2 Mirror | 2 Mirror | 2 Mirror | 2 Mirror | 2 Mirror |
| 3 Lamp | 3 Lamp | 3 Clock | 3 Lamp | 3 Clock |
| 4 Chess | 4 Noticeboard | 4 Shelves | 3 Chess | 4 Lamp |
| 4 Noticeboard | 5 Chess | 4 Chess | 4 Noticeboard | 5 Noticeboard |
| 4 Clock | 5 Table | 5 Lamp | 4 Table | 6 Chess |
| 5 Table | 5 Penpot | 5 Noticeboard | 5 Clock | 7 Table |
| 6 Penpot | 5 Shelves | 6 Table | 6 Penpot | 8 Shelves |
| 7 Shelves | 6 Clock | 7 Penpot | 6 Shelves | 9 Penpot |

**Table four: overall ranking**

We split the responses into three bins 0-1: dislike, 2: neutral, 3-4: like and show the breakdown for each solution for all respondents (Figure 4).

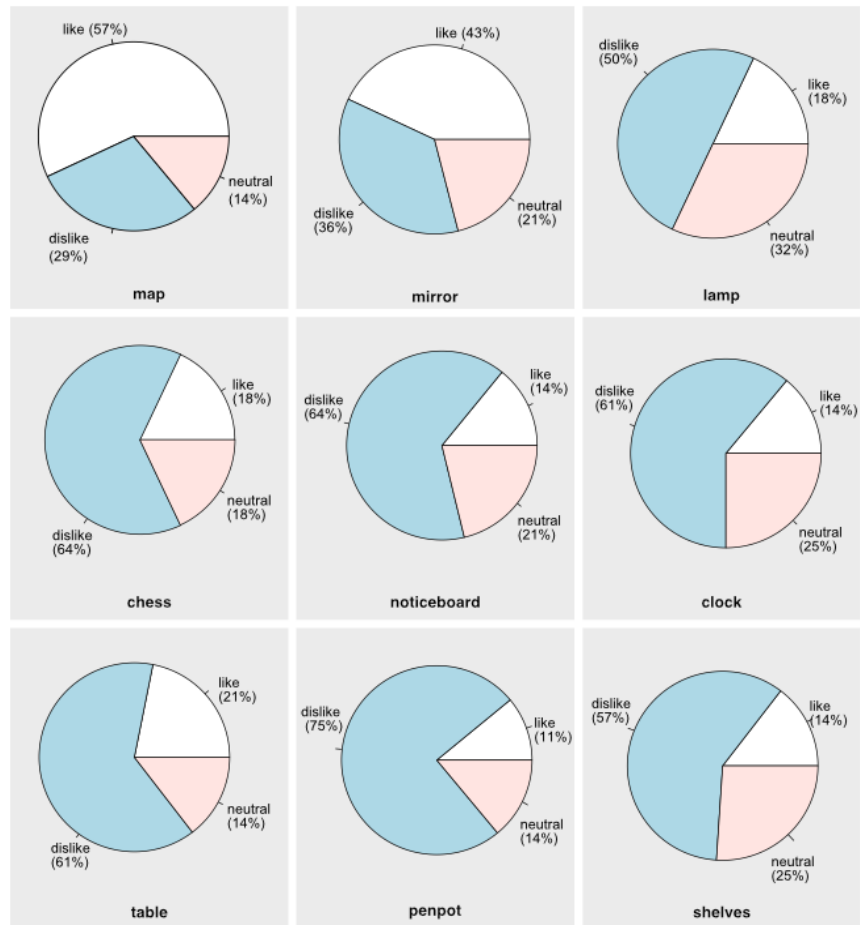

**Figure 4: users' reactions (dislike,neutral,like)**

The majority of users like the map and mirror (Figure 1). Half or more of the respondents disliked the remaining solutions and the penpot was the most disliked; 75% of respondents provided a score of 0 or 1 (dislike).

Our respondents were asked to rank each of the solutions in terms of their perceived security on a three point scale (insecure (0), mostly secure(1), secure(2)). In spite of a positive overall response, the wall map was deemed to be less secure than most of the other solutions. Table four shows the sum of security rankings for all users and the rankings for experts vs non-experts. There is agreement on the most secure solution (mirror), and the penpot (most disliked overall) and chessboard are both ranked in the top three for each. This gives us our first design observation.

***Design observation one: Perception of security may not be the overriding factor when it comes to a user's acceptance of a security solution.***

| Solution | Security ranking (all) | Solution | Security ranking (good/pro) | Solution | Security ranking (little/none) |
|---|---|---|---|---|---|
| mirror | 35 | mirror | 16 | mirror | 19 |
| chess | 33 | chess | 15 | penpot | 19 |
| penpot | 33 | penpot | 14 | table | 18 |
| noticeboard | 31 | noticeboard | 14 | chess | 18 |
| clock | 30 | clock | 14 | shelves | 17 |
| lamp | 29 | lamp | 13 | noticeboard | 17 |
| map | 28 | map | 13 | clock | 16 |
| table | 28 | shelves | 11 | lamp | 16 |
| shelves | 18 | table | 10 | map | 15 |

**Table five: security ranking**

In relation to memorising secrets, we asked our respondents to assume the solutions would be used to encode multiple secrets/combinations. We asked them to rank how well these secrets would lend themselves to being remembered (0, very difficult – 4, very easy). The mirror and map are ranked most highly, with the shelves performing most poorly.

| Solution | Memory ranking (sum of rankings) |
|---|---|
| mirror | 83 |
| map | 60 |
| table | 46 |
| lamp | 43 |
| chess | 40 |
| penpot | 33 |
| clock | 30 |
| noticeboard | 28 |
| shelves | 17 |

**Table six: ease with which secrets can be memorised**

## V.    DISCUSSION ON FEEDBACK

In this section we attempt to provide further context for the results we've presented so far. In our questionnaire we asked users to (optionally) provide further feedback on the scores that they provided as well as to reflect more broadly on each solution. We discuss three universal concerns: usability, practicality and security as well as two solution-specific concerns: trust/privacy and cost. As we discuss the feedback we highlight a set of design challenges and constraints that, whilst not peculiar to this class of solution, are a intended as a useful reference point for future development for this type of MFA solution. We conclude with a brief discussion on a set of general observations that provide further context for developing solutions in this space.

**Usability**   The set of pictures and descriptions of the solutions were sufficient for respondents to imagine using them within their homes. Respondents critiqued the solutions in relation to the effort required to use them. Some respondents mentioned that "*The novelty would soon wear off.*" and that "*it would annoy me after not very long*". Others spoke of a perceived effort in use: "*this seems like a lot of work just to authenticate*". Some solutions, such as the noticeboard were seen by some to be clumsy or difficult to use: "*Fiddly and feels like a pain to have to reorganise all this stuff.*"

By embedding a secondary purpose in household objects, however, some respondents suggested that the interactions could elevate a purely functional, mundane practice into something enjoyable: *'It feels like fun - and I like the clock!'* and *"it would make me feel like a spy :)"* and (or the maps) *feels like a more enjoyable thing that you can potentially create a narrative for"* and *"I love maps. I love putting pins in maps. I love that this is exactly like an escape room."* Furthermore, some respondents proposed additional lifestyle benefits such as encouraging better organisation *("It could have the side-effect of breaking bad organisational habits around the home I guess.")* and managing access to consoles: *"Alternatively, with a few design tweaks it could lock them out of their x-boxes and encourage them to pick up a book to read!"*

***Design observation two:  Though important to keep interactions simple, some complexity might be offset by developing playful, personal or even independently useful interactions around a solution's primary purpose.***

**Practicality.** This relates to feedback on the choice of objects to repurpose and the perception of pragmatic influences of the household (i.e. routines, lifestyle, inhabitants, environment) on usage. Several subthemes emerged:

> *Aesthetics.* Unsurprisingly, the 'look' of an object and how well it would fit within a home was a commonly expressed view, from simply not wanting a particular artefact: *"Would not necessarily want a large map on the wall.",* to concern about style or design: *"May end up being more about the aesthetics of how it looks than what it does."* and *"I don't want the style of my home decor dictated by the need to accommodate security devices / home furnishings to have markings that spoil their aesthetics.",* to consideration on how it would fit within a particular space*: "'Feels less dependent on a recognisable auth object, and so would more easily blend into my home".*
>
> *Fit within home.* Somewhat related to aesthetics is whether an object would have a place within a user's home. Some respondents outright rejected a solution simply because the primary object had no obvious place: (*"I do not have a place in my home where an angle poise lamp is practical"*) or represented clutter (*"So I need another pot to store the spare pens?"* / *"most people already have clocks so are unlikely to need to be purchasing one anyway."* / *" I like the idea albeit that it would require a specific purchase."*)  or not common enough to most households *("Not everyone has world maps"* / *"one is unlikely to purchase a chess set for the specific purpose of security."* / *"the object may not be a desirable purchase for the end-user.").*  Conversely, the advantage of leveraging more commonly available objects was noted *"Tried to think of an alternative system, but found it difficult. Everyone has books on shelves.".*  There were, also concerns about placement and access :  *"if I need to be in this part of the house to do this thing then I'm not happy."* and *"It also ties me to one place for authentication"* and *"I want to lay down on my couch and not move to put random books in a random position which I will have to put back again later".*  Finally there were a range of concerns related to accessibility: *"If the clock is on the wall you'd have to climb on a chair."* and *"Would the colour work for colorblind people? And in the dark?"*
>
> *Fit with lifestyle.* Respondents also reflected upon how their lifestyle/situation would influence and be influenced by using a particular artefact.  The influence of other householders, for example, came strongly to bear, i.e. the relative inaccessibility of the clock on the wall was deemed an advantage in a household with children: *"Given its location it is less likely to suffer from interference, e.g. by small children.",* but there were other concerns around the ability for other members to (accidentally, unknowingly) interfere with the solution: "*What if someone "borrowed" one of the pens you needed.*"   Specific aspects of a user's lifestyle or situation could also render a solution impractical "*It wouldn't fit with my lifestyle - I have too many bookshelves and books wander around the house...I may not be able to locate the right books at the right time!*" and "*The approach makes too many assumptions that are impractical in daily life. When sharing a household it is possible for things to be moved, books to be used and layouts to be forgotten.*" In the case of the chess board, lifestyle (i.e. whether users play chess) was had the predominant influence on a solutions acceptability one user ranked the chess highly simply stating in the feedback that "*I play chess regularly*", whereas others stated "*I don't play chess, so board configurations don't have meaning for me.*"

***Design observation three:  The primary purpose of an artefact bears as much or even more influence on a solutions acceptability than how it has been repurposed.***

***Design observation four: Lifestyle artefacts (i.e. chess board, map) will appeal more greatly to a smaller userbase than utility artefacts (e.g.: lamps, shelves).***

*Tension with primary purpose.* Respondents noted cases where the authentication functionality was: *i.* at odds with, *ii.* interfered with the primary purpose of the object or *iii.* was prone to failure under specific situations; this is especially the case with utility objects. The following feedback for the lamp, noticeboard, table and shelves illustrate the point:

| lamp | table | shelves | noticeboard | penpot |
|---|---|---|---|---|
| "*I use it to see. If i'm also using it as a security system i'm preventing myself from using it to see*"<br><br>"*would be messing up the lighting set up every time I used it. It also begs the question, how well would it work in the daytime?*".<br><br>"*I don't like needing to reposition my lamp all the time. Once positioned I rarely need to move it*" | "*You can't use the table for other things / it would have to be cleared of other items in order to do a security check.*".<br><br>*Tables have a certain logical and typical layout (laptop is turned toward the seated user for example).* | *I don't want to store my books based on a security code pattern, I want to store them in a way that I determine so I can find a book when I want it.*<br><br>*What happens if a book is taken for its original purpose (to be read) by another person* | *Also makes the board difficult to use for actual utility.*<br><br>*I'm not moving the pins on the notice board, they're here for a reason.* | *Why would I want a pen pot that I can't use as a pen pot?* |

**Table seven: tension with primary purpose**

***Design observation five: The method of authentication must be orthogonal to an artefact's primary purpose.***

**Solution specific concerns**

In addition to the concerns and design challenges raised across the whole set of solutions, a few concerns were raised that were unique to one or a few artefacts:

**Cost**

Cost was only considered two times as a concern for the mirror: "*concerns would be cost*" and the atomic clock "*are you planning to give me an atomic clock? Looks rather expensive*" so is perhaps more interesting for its lack of recurrence as a theme.

**Trust/Privacy**

Trust and privacy were considered by four respondents in relation to the mirror. One respondent had a general wariness: "*I would feel like the mirror was watching me, it would fully freak me out.*", whereas others cite specific concerns with the use of a camera "*Is this a mirror or a disguised camera??* and *"I assume the mirror uses a camera for item identification?"* and with it, the potential to violate privacy: "*What's it looking at the rest of the time?"* and "*If the photo of the item includes background / the room behind it - this is an intrusion on privacy*".

**Security**

In this section we take a deeper dive into participants' feedback on the perceived security of our solutions. We asked participants to evaluate whether they believed the solution to be: *insecure*, *mostly secure* or *secure* and to provide further information on the reasons for their answer. We did not specify a particular threat model (beyond stating the setting was householders at home), leaving it to participants to elaborate as necessary. We have broken down the feedback into five broad concerns; *choice and use of secrets*, *insider threats*, *timing* and *robustness*.

**Choice and use of secrets**

As with any system that relies on secrets there is a tension between the ease with which secrets can be memorised and the vulnerability to exposure. We have further broken the feedback down into the size of the secret space, secret choice and secret management (i.e. how it is handled by the system).

> *Size of secret space*: This concerns the number of potential combinations and entropy in a system (and by implication, susceptibility to brute force attacks). All solutions were evaluated on this basis by at least one respondent. Participants distinguished between the *theoretical* and *practical* size of the space. The clock and lamp were most commented on in relation to (small) theoretical secret space ("*Limited number of possible configurations*" (clock) / "*There seems to be a smaller number of combinations than the other methods*" (clock) / "*Just not sure how many combinations there are.*" (clock) / "*I like this a bit, but it's not clear how big the solution space is (number of possible lamp positions might be limited) thus susceptible to brute force attack*" (lamp) / *Assuming six possible positions for each pivot* *point (3^6), it is slightly more vulnerable to a brute force attack than a simpler combination lock with 10^3 combinations.* (lamp) / *You could just guess the configuration or do it by mistake* (lamp). Unsurprisingly the chess set was viewed as having a sufficiently large theoretical combination space *("Huge number of configurations combined give good level of security"* (chess)). Even in cases where there is no theoretical limit on the number of possible combinations (given that many objects may be used in authentication) there was still concern that practically, the secret space could be too small i.e. users would only make use so many books, pens or objects. *("depending on the number of books, attempts could be made to crack the order"* (shelves) / *"People could randomly put the books in the right order"* (shelves) / *"You could just put a random selection of pens in to guess - again people would keep the pens close by so it would potentially be quite easy"* (pens))

**Design observation six:  *A theoretically large secret space can be significantly shrunk by practical constraints (such as availability and use of authenticating objects, position and location of object).***

*Secret choice:*  We asked participants to comment on how easy they thought it would be to memorise sets of combinations if using devices for multiple authentication purposes (*"Assuming you used this approach for a few different tasks, how easy do you think it would be to remember different combinations?"*).  We have already noted that the map and mirror ranked most highly.  Where participants could envisage a model ("*narrative*") that would help remember combinations, they reacted favourably.  The world map scored better than most because it presents an obvious technique ("*Easier to remember as can remember places rather than shapes"* / *"I can imagine remembering a list of countries into which to stab my pins more easily than some of the other less mnemonic-friendly examples"* / *"Place names are more memorable combination for me"*).  The mirror scored highly simply because it doesn't require combinations, but single recognisable/memorable objects (*Just need to remember one object / I love this, as it's essentially arbitrary object as password*)**.** For other solutions, responses were clearly influenced by whether participants were able to impose their own memory system on the solution ("*Could be arranged logically (e.g: keys on the doorside of the board) without that decision being overtly apparent*" (noticeboard) / *this is similar to the mirror object example, where arbitrary objects are my password, but with the addition of ordering. I like it.* (table).  In most cases, in the absence of an obvious approach, participants felt that memorising secrets would be challenging ("*I think remembering the grid positions in this example is much harder. I'm essentially having to memorise several sets of coordinates" (*noticeboard*)* / "*I can imagine struggling to remember the necessary order of books for multiple tasks*" (shelves) / *Pens rarely stick in my mind. Combinations of pens less so.* (penpot).  The chess board, as with most of its feedback, was positively received by chess players (*fairly straightforward to create mental rules to memorise positions and reproduce them quickly)* and not by non-players (*I am not a chess master. I would never remember the config so would have all sorts of offline ways to remember the config, introducing flaws*).  In the absence of an obvious memory technique, participants simply assessed the solution on how easy the raw information would be to memorise; with the atomic clock, for example, which embeds secrets in the combinations of coloured balls around its circumference, one participant stated: "*the atoms here give a relatively easy way to remember the system*" whereas another stated *"can literally anyone in the world remember a colour sequence?".*  In cases where memory is seen as too challenging, respondents raised concerns that users would record secrets in some way.  (*Really un-user friendly in term of remembering configurations - would need photographs. / The myriad of possible layouts means that some form of written* *or photographic record is likely to be kept* (chess) / *can see a lot of people leaving them in the right configuration and almost definitely writing them down (chess) /*

In tension with the ease with which secrets can be remembered is the choice of a suitably strong secret. The chess board, for example was considered by a few respondents to be vulnerable to predictable choices, such as standard opening moves (*people would choose actual scenarios and these are widely available / well say it was the Sicilian defence, that config is widely available)*. Other concerns were raised about the potential for personal information, known to others, being used in secrets that could then be compromised (we cover this in more detail later).

***Design observation seven: Users prefer systems where the patterns or combinations of the secret in an object can be aligned to a narrative, so long as not easily guessed by others.***

**Secret management.** Respondents identified other concerns around the ongoing use and management of the secrets. One concern, raised in different ways, related to human fallibility and the effort required to keep secrets secret. After authentication for example, when a secret has been accepted, it will remain visible others until the system is transitioned into an unrevealing state. Respondents mentioned that users may simply forget to move them out of these states ("*Forgetting to move spheres after authentication."* (clock) / *"I can see a lot of people leaving it in a particular position"* (lamp) / *Forgetting to move the pins out of the correct configuration after use.* (map)/ *The homeowner/authorised user might forget to replace the pegs after authentication* (noticeboard) / *forgetting to move the books again*), but also that there is inherent difficulty in transitioning to a neutral/random state for some solutions (*"on my desk it is not empty and would be hard to empty, thus it is likely to stay in the same configuration after the verification"* (table) / *I do, however, think it suffers from the danger of users having the correct configuration as their default setup and changing only a single pin.*" (map) / *"if I leave my chessboard in a particular layout, I'm offering up my password." (chess*) / *"high effort so people don't bother"* (shelves). The penpot, at least, was seen to be much easier to transition (*clever, easy to scramble*) and the mirror was seen by some respondents as better ("*Nothing persistent to scramble again afterwards"* / *"Because the setup is evanescent, there's no risk of leaving it set up."),* respondents were concerned that situations could still arise where hints are inadvertently revealed ("*If the system is used regularly then the objects are likely to be lying around near the mirror so easy to guess"* / *"You could just guess objects - most people would keep them near to the mirror"* / *"The key will be the item next closest to the mirror"* / *"Would probably leave all the objects by the mirror"* ).

***Design observation eight: It is as important to make secrets as easy to set as it is to unset. Solutions must account for fallibility (i.e., inadvertent exposure or hints remaining after use).***

**Insider threats** All of the solutions we have considered are expected to operate in multi-occupancy households. A primary security threat emerges from other members who may have access to the authenticating device and know personal information about its owner. Several concerns were raised by respondents:

*Conspicuousness of tool 'out of placeness'*: Assuming some reliance on our solutions being 'hidden in plain sight", some objects were viewed as being incongruous to either their surroundings or the household inhabitants ("*Would stand out in a non-chess player's home"* (chess) / *"Table would be left obviously clear"* (table) / *"It would likely be recognised by anyone trying to hack into my account as being the authentication tool"* (map)

*Observability:* Three different threats were raised in relation to observability. First there were concerns around *shoulder surfing*: householders being able to view the use of solutions during and after authentication (*"visibility of config in public place might be compromised by another user viewing the pegs, which would presumably stay in place post authorisation."* (noticeboard) / *"[inadvertent display of] the correct order or position of books for authentication to others present who are not allowed to see it"* (shelves) / *"the lamp needs to be pivoted in secrecy"* (lamp). Second, there were concerns that images that reveal configurations could be inadvertently captured and shared on social media (*"A picture of the set positions being taken inadvertently/on purpose and being shared on social media."* (chess) / *"As with others there is a risk of the clock and layout appearing in domestic photos."* / *"otherwise a picture of the clock in the background of a selfie could weaken the security"*). Third, there was concern, particularly with the map, but also with the noticeboard and clock that continued use and wear and tear would reveal secrets over time (*"Pins leaving traces/dust etc., giving clues to settings."* (map) / *"Pinholes in the map indicate previously placed pins."* (map) / *"Pin positions already used would be permanently recorded"* (map) / *"Holes will remain and show frequent positions"* (map) / *"Perhaps it wouldn't be too difficult to read where pinholes have been on the map)."* (map) / *"Nobbles could also be left in the 'authenticate' position unless they shuffle for each challenge."* (clock) / *"Those grids become discoloured, so we're opening ourselves to a limited set of options to brute force"* (noticeboard)

*Tampering:*. This concerns both deliberate and (perhaps more commonly) accidental tampering with the solution by others. There is some tension between the solution keeping its secondary purpose secret, to reduce the insider threat, and limiting the likelihood of visitors and householders from accidentally thwarting its use. These concerns were explicitly raised with the lamp, noticeboard, shelves, table and pens, though are clearly applicable to the other solutions (*"Depending on where the lamp is placed, the homeowner might not be the only person with opportunities to operate the lamp"* (lamp), *"what if someone takes a book out of the shelf?"* (shelves), *"very good chance that the pegs would be lost by other members of the household"* (noticeboard), *"anyone can move objects"* (table))

*Knowledge of person:* This concern may be viewed as an extension of the concerns around choice of secret. It was specifically raised in relation to the map; that it is probable that other household members would know countries and regions that are meaningful to others and could exploit this to guess combinations (*"It's better than the lamp/books because it might be personally meaningful and easier to remember, however, that also raises a security weakness."*(map) / *"This is interesting as the pegs could be arranged without relying on personal history such as with the map."* (map) / *"If you knew something about the user, it might inform a reasonable guess, eg place of last holiday."* (map) / *"Friends knowing the countries I like/have visited on specific trips."* (map)

**Design observation nine: Users both care about and readily anticipate the influence of other householders on a proposed solution, particularly regarding accidental or deliberate tampering.**

**Timing**

Several respondents raised the issue of timing, i.e. whether the solutions are used once during authorisation or must remain in place for the entire time a privileged action is taking place. As one respondent put it, regarding the shelves: *"As for books (and indeed all options) there seem to be risks in determining how long the required configuration lasts - too short and you may lock out the authorised user; too long and you may admit the unauthorised user"*. Others wanted similar clarification: *"You haven't said whether the book configuration would change each time, or how long it lasts [shelves]"*. Additionally, there is a critical issue that emerges from the dual nature of the solution which was touched upon in part with some of the feedback on secret choice, but was made more explicit by one respondent: *"Little/no scope for locking down system if dual purpose; cannot know if auth attempt or just normal use."* That is, solutions must be able to distinguish between purposes of use to prevent accidental authentication and brute force attacks.

***Design observation ten:  To prevent brute-force or accidental authentication, a solution must know when a user's interactions are for the primary (i.e., original) or secondary (i.e., authenticating) purpose.***

**Robustness**

Users expressed concerns about system failure; most common was the concern that critical components (i.e. tokens) might be lost, damaged, stolen or borrowed and prevent users from authenticating: "*what happens if the specific items is lost/thrown away because they're damaged? Also I WILL lose the tiny balls, immediately, no doubt in my mind at all. [clock]" / Have you ever tried to find a pen in your house when you need one? They are never where they are supposed be are they?* [penpot] / *"I'm going to lose pens and be locked out of my vpn, Objects likely to walk in busy family household"*[penpot*] / ,Relies on me keeping all that stuff in my house and remembering to not throw something out that I might need to authenticate* [noticeboard] / 'i am a loser, i lose things.' [noticeboard]

***Design observation eleven: Users do not like systems that utilise 'unfixed' items (e.g. pens, pegs) as authentication tokens, fearing their loss or destruction will result in loss of access.***

## VI.　　CONCLUSION

In this paper we have explored a tangible key exchange mechanism for supporting secure peer-to-peer communications. We argue that by making *physical* in-home interactions a requirement of key exchange (*what you have, where you are, what you know*), we can mitigate a whole class of online attacks. To assess the viability of the overall solution space, and to understand the features of a system that are important to gaining acceptability, we have sketched nine different solutions, that: *i.* are technically feasible *ii.* utilise tangible interactions with domestic fixtures and fittings, *iii* are constrained by location and *iv.* utilise 'hidden-in plain-sight' secrets. We ran an online experiment that presented a set of simple use cases for each of these solutions to a mix of security professionals and lay people. We gathered feedback on the overall concept, the acceptability (i.e., whether users would consider using a solution) and the perceived security. We found that approximately 75% of users would use at least one of the presented solutions and that the variability in acceptance was driven by a rich set of concerns and observations. We have highlighted eleven key observations that will provide a useful contribution to further exploration and development of tangible, in-home key exchange mechanisms. These are as follows:

- Perception of security may not be the overriding factor when it comes to a user's acceptance of a security solution.

- Though important to keep interactions simple, some complexity might be offset by developing playful, personal, or even independently useful interactions around a solution's primary purpose.
- The primary purpose of an artefact bears as much or even more influence on a solution's acceptability than how it has been repurposed.
- Lifestyle artefacts (i.e., chess board, map) will appeal more greatly to a smaller userbase than utility artefacts (e.g.: lamps, shelves).
- The method of authentication must be orthogonal to an artefact's primary purpose.
- A theoretically large secret space can be significantly shrunk by practical constraints (such as availability and use of authenticating objects or the position and location of object).
- Users prefer systems where the patterns or combinations of the secret in an object can be aligned to a narrative, so long as it's not easily guessed by others.
- It is as important to make secrets as easy to set as it is to unset.  Solutions must account for fallibility (i.e., inadvertent exposure or hints remaining after use).
- Users both care about and readily anticipate the influence of other householders on a proposed solution, particularly regarding accidental or deliberate tampering.
- To prevent brute-force or accidental authentication, a solution must know when a user's interactions are for the primary (i.e., original) or secondary (i.e., authenticating) purpose.
- Users do not like systems that utilise 'unfixed' items (e.g., pens, pegs) as authentication tokens, fearing their loss or destruction will result in loss of access.

Our works suggests that *physical* in-home interactions offer a promising (and much needed) design space for secure, usable key exchange.  User acceptance will be subject to a set of concerns that extend beyond security and usability to include aesthetics, domestic arrangements and personal routines and interests.

## ACKNOWLEDGEMENTS

This work was supported by the PETRAS 2: EP/S035362/1 National Centre of Excellence for IoT Systems Cybersecurity under the project "Tangible Security" (TanSec).

MARCH 2022                                                                                                                              20**(Rajivan, 2017)**  P. Rajivan, P. Moriano, T. Kelley, and L. J. Camp, "Factors in an end user security expertise instrument," Information Computer Security, vol. 25, no. 2, pp. 190–205, 2017.

**(Reynolds, 2018)**       J. Reynolds, T. Smith, K. Reese, L. Dickinson, S. Ruoti, and K. Seamons, "A tale of two studies: The best and worst of yubikey usability," in 2018 IEEE Symposium on Security and Privacy (SP), pp. 872–888, IEEE, 2018.

**(Stanton, 2005)**J. M. Stanton, K. R. Stam, P. Mastrangelo, and J. Jolton, "Analysis of end user security behaviors," Computers Security, vol. 24, no. 2, pp. 124 – 133, 2005.

**(Stajano, 2011)** Stajano, F., 2011. Pico: No more passwords! In: International Workshop on Security Protocols. Springer, pp. 49–81.

**(Ometov, 2019)**  A. Ometov, V. Petrov, S. Bezzateev, S. Andreev, Y. Koucheryavy and M. Gerla, "Challenges of Multi-Factor Authentication for Securing Advanced IoT Applications," in IEEE Network, vol. 33, no. 2, pp. 82-88, March/April 2019

**(Sinigaglia, 2020)** Federico Sinigaglia, Roberto Carbone, Gabriele Costa, Nicola Zannone,A survey on multi-factor authentication for online banking in the wild,Computers & Security,Volume 95,2020,101745,ISSN 0167-4048

**(Shrestha, 2018)** Shrestha, P., Saxena, N., 2018. Listening watch: Wearable two-factor authentication using speech signals resilient to near-far attacks. In: Proceedings of the 11th ACM Conference on Security & Privacy in Wireless and Mobile Networks. pp. 99–110.

**(Reynolds, 2020)** Reynolds, Joshua, et al. "Empirical Measurement of Systemic 2FA Usability." 29th USENIX Security Symposium (USENIX Security 20). 2020.

**(Varshavsky, 2007)** Varshavsky, A., Scannell, A., LaMarca, A. and De Lara, E., 2007, September. Amigo: Proximity-based authentication of mobile devices. In *International Conference on Ubiquitous Computing* (pp. 253-270). Springer, Berlin, Heidelberg.

**(Y. Xie)** Y. Xie, C. Shi, Z. Li, J. Liu, Y. Chen and B. Yuan, "Real-Time, Universal, and Robust Adversarial Attacks Against Speaker Recognition Systems," *ICASSP 2020 - 2020 IEEE International Conference on Acoustics, Speech and Signal Processing (ICASSP)*, Barcelona, Spain, 2020, pp. 1738-1742, doi: 10.1109/ICASSP40776.2020.9053747.

**(Zhang, 2012)** Zhang, F., Kondoro, A. and Muftic, S., 2012, June. Location-based authentication and authorization using smart phones. In *2012 IEEE 11th International Conference on Trust, Security and Privacy in Computing and Communications* (pp. 1285-1292). IEEE.